\documentclass[twocolumn,english,showpacs,amsmath,amssymb,aps,prb, floats,groupedaddress,superscriptaddress]{revtex4}
\usepackage[]{fontenc}
\usepackage[latin1]{inputenc}
\usepackage{subfigure,amsmath,epsfig,dcolumn}
\usepackage{bm,amsfonts,graphicx,babel}

\newcommand{\bra}[1]{\langle #1 |}
\newcommand{\ket}[1]{| #1 \rangle}
\newcommand{\be}{\begin{equation}}
\newcommand{\ee}{\end{equation}}
\newcommand{\beq}{\begin{eqnarray}}
\newcommand{\eeq}{\end{eqnarray}}
\makeatother

\begin{document}

\title{Nanoscale phase separation and superconductivity in the one-dimensional Hirsch model}

\author{Alberto Anfossi}
\affiliation{Dipartimento di Fisica dell'Universit\`a di Bologna, viale Berti-Pichat 6/2, I-40127, Bologna, Italy}
\author{Cristian Degli Esposti Boschi}
\affiliation{CNR, Unit\`a di Ricerca CNISM di Bologna, viale Berti-Pichat 6/2, I-40127, Bologna, Italy}
\affiliation{Dipartimento di Fisica dell'Universit\`a di Bologna, viale Berti-Pichat 6/2, I-40127, Bologna, Italy}
\author{Arianna Montorsi}
\affiliation{Dipartimento di Fisica del Politecnico, corso Duca
degli Abruzzi 24, I-10129, Torino, Italy}

\pacs{05.30.Fk, 71.10.Fd, 71.10.Hf}

\date{January 26, 2009}

\begin{abstract}
We investigate numerically at various fillings the ground state of the one-dimensional Hubbard model with correlated hopping $x$ (Hirsch model). It is found that, for a large range of filling values $n$ around half filling, and for repulsive Coulomb interaction $u\leq u_c(x,n)$, phase separation at a nanoscale (NPS phase) between two conducting phases at different densities occurs when $x\gtrsim 2/3$. The NPS phase is accompanied by the opening of a spin gap and the system behaves as a Luther-Emery liquid with dominant superconducting correlations. Close to half filling, an anomalous peak emerges in the charge structure factor related to the density of doubly occupied sites, which determines the size of the droplets in the NPS phase. For $1/2\lesssim x\lesssim 2/3$ a crossover to a homogeneous phase, still superconducting, takes place.
 \end{abstract}

\maketitle

\section{Introduction}\label{secI}

The subject of phase separation (PS) in strongly correlated fermionic materials has been widely investigated in connection to various physical systems, ranging from high $T_c$ materials \cite{HTC-PS} to cold fermionic atoms trapped in optical lattices.\cite{CF-PS} In particular, it has been noticed how PS often occurs close to the transition to a superconducting (SC) behavior. There is still no full explanation of such an observation.  Since in high--$T_c$ materials the SC phase appears upon doping an antiferromagnetic insulator, in the past the attention of both experimentalists and theoreticians has been mainly focused on PS in the underdoped regime, between an undoped insulator and a doped metal.  More recently, experimental compelling evidence has grown of the presence of two types of charge carriers\cite{PS_SC} in cuprate superconductors and the occurrence of nanoscopic phase separation of the two components has been investigated (see for instance Ref. \onlinecite{PS_NANO} and references therein).

On the side of microscopic theoretical modeling, correlated electronic materials are well described by the Hubbard model and its extensions. In this context, PS mainly appears as segregation of an insulating phase (either the half-filled antiferromagnet or the immobile pairs) within a low-density conducting phase.\cite{PS-tJ,PEMI,CSC} Very recently \cite{MONT} it was shown that in some cases PS can occur as well as coexistence of two conducting phases. In particular, a high-density conducting phase implies the presence of mobile pairs in the system. In this paper we explore the possibility that phase coexistence of two phases of different densities in the charge degrees of freedom generates the appearance of SC order, when accompanied by the opening of a spin gap. In fact, a nonvanishing spin gap may induce phase separated droplets of nanoscale size in the systems, and subsequent  quasi-long-range correlations between the mobile pairs of the different droplets, allowing the superconducting correlations to become dominant.\\
The model we deal with is the one-dimensional Hubbard model with correlated hopping, the latter describing the interaction between charges located on bonds and on lattice sites. The model's Hamiltonian reads:
\be
    H_{BC} =  -\sum_{\langle ij\rangle, \sigma} \left [1-x(n_{i\bar\sigma} +n_{j\bar\sigma})\right ] c_{i\sigma}^{\dagger}c_{j\sigma} + u\sum_{i} n_{i\uparrow}n_{i\downarrow}
    \; , \label{ham_BC}
\ee
where $c^\dagger_{i\sigma}$ creates a fermion with spin $\sigma=\{\uparrow,\downarrow\}$, $\bar\sigma$ denoting the opposite of $\sigma$, $n_{i\sigma}=c^\dagger_{i\sigma} c_{i\sigma}$ is the $\sigma$-electron charge, and $\langle ij \rangle$ stands for nearest-neighboring sites. The parameters $u$ and $x$ are the Coulomb repulsion and the bond-charge interaction amplitude, respectively, and the lower case symbols denote that the coefficients have been normalized in units of the hopping amplitude. Moreover, $N$ is the number of electrons on the $D$-dimensional $L$-sites lattice, so that $n=N/L$ is the average filling value per site. The model has been widely studied in the literature (see for instance Ref. \onlinecite{HBC} and references therein, as well as Refs. \onlinecite{hirsch} -\onlinecite{DUAN}). In particular, since $H_{BC}$ is not invariant under particle-hole transform, it has been proposed in two dimensions by Hirsch motivated by a theory of hole superconductivity.\cite{hirsch}

In $D=1$, it has by now become clear that the basic features of the model are well captured in the weak-coupling limit ($x \ll 1$) by the standard bosonization approach,\cite{JAMU} and resemble those of the ordinary Hubbard model with rescaled interaction; while for $x\gtrsim{1\over 2}$ a quite different scenario emerges at half filling.\cite{ADMO,AAA} In particular, for not too large on-site Coulomb repulsion $u< u_c(x)$ a new phase appears, characterized by slowly decaying SC correlations, and incommensurate fluctuations in the charge correlations: the phase was denoted as ICSS (incommensurate singlet superconductor). The physical origin of such a change in the system has not yet been fully explained, though a subsequent paper\cite{MONT} suggested it could be related to the presence of PS between conducting phases with different Fermi momenta.

Here we shall show --through a detailed density-matrix renormalization group (DMRG) numerical study-- that in fact the SC phase found at $n=1$ for $x\gtrsim 1/2$ survives for a wide range of filling values. Moreover, for $x\gtrsim 2/3$, it turns out to be characterized by nanoscale phase separation (NPS).

In particular, in Sec. \ref{secII} we first discuss some aspects of the model in $D=1$ and its known properties. We then proceed to the study of the filling dependence of the ground-state energy and chemical potential (Sec. \ref{secIII}), and of pair-pair and charge-charge correlations and spin gap (Sec. \ref{secIV}). Finally, we investigate in Sec. \ref{secV} the charge structure factor, thus obtaining an estimate of the Luttinger exponent $K_\rho$ that controls the decay of charge and pair correlators. The discussion of the results and some perspectives are given in Sec. \ref{secVI}.

\section{Hirsch model in $D=1$}\label{secII}

The bond-charge Hamiltonian (\ref{ham_BC}) has both spin [$su(2)$] and charge [$u(1)$] symmetries. It is not invariant under particle-hole transform, though the latter can be implemented to show that the range $0\leq x\leq 1$ is in fact representative of the behavior of the model at any  $x$ value. Indeed, under the transformation $c_{j\sigma}\rightarrow [\mbox{sgn}(2 x-1)]^j c_{j\sigma}^\dagger$, $H_{BC}$ transforms (up to constants) as
\be
   H_{BC} (x,u) \longrightarrow |2x-1| H_{BC}\left( \mbox{sgn}(2 x-1) \bar x, \bar u \right) \; ,
\ee
where $\bar a= \frac{a}{|1-2 x|}$ and $a=x,u$. Hence, the interval $0\leq x\leq 1/2$ can be mapped into $0\geq x\geq -\infty$, the Hubbard model ($x=0$) being representative of this regime, while the complementary range $1/2\leq x\leq 1$ can be mapped into $1\leq x\leq +\infty$ and one expects that the integrable case $x=1$ (Ref. \onlinecite{AAS}) is representative of such a different regime. In this respect, the point $x=1/2$ assumes a special role in that it can be simultaneously mapped into the two limiting cases $+\infty$ and $-\infty$.

\begin{figure}
\includegraphics[width=80mm,keepaspectratio,clip]{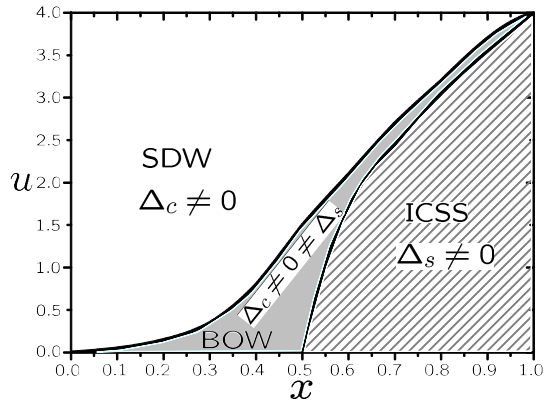}
\caption{$T=0$ phase diagram of the one-dimensional Hirsch model at half filling (Ref. \onlinecite{AAA}). In each phase the gapped sectors (charge, $\Delta_c$ and/or spin $\Delta_s$) are reported; SDW stands for spin-density waves, while BOW stands for bond ordered wave.} \label{fig1}
\end{figure}

In $D=1$, as far as $x\ll 1$, the Hirsch model can be approached by means of  the bosonization technique.\cite{JAMU} It was found that Eq. (\ref{ham_BC}) corresponds to an effective Hubbard model with rescaled Coulomb interaction $u_{eff}$ and Luttinger parameter $K_\rho$ given by
\be
    u_{eff}= \frac{u+8 x \cos k_F}{1-n x} \, , \quad K_\rho=\frac{1}{\sqrt{1+\frac{u_{eff}}{v_F}}}\; , \label{ubos}
\ee
where $k_F= n {\pi\over 2}$ and $v_F$ is the Fermi velocity. The result implies that the metal-insulator transition at half filling still occurs at $u=0$. At the same time, it suggests that for $n>1$, $u_{eff}$ may actually become negative for sufficiently small $u>0$, so that a  spin gap opens, and the system is expected to enter a Luther-Emery liquid (LEL) phase. Since in this case also $K_\rho>1$, the LEL phase should have dominant SC correlations. Numerical investigations --both at $T=0$ (Refs. \onlinecite{AAG} and \onlinecite{QSZ}) and at $T\neq 0$ (Ref. \onlinecite{KESC})-- confirmed the validity of this scenario for $n>1$, in some cases also at larger $x$-values.\\
On the other hand, at $x=1$ the model acquires extra symmetries and both the thermodynamics \cite{DOMO} and the $T=0$ phase diagram\cite{AAS} can be obtained. The latter turns out to differ from that of the Hubbard model in many aspects. First of all, at half filling the metal-insulator transition moves up to $u=4$. Furthermore, below such value of $u$ a new phase characterized by the presence of pairs and off-diagonal long-range order appears. This phase survives for a large range of filling values around $n=1$. Since the model in this case has no spin gap $\Delta_s$, the phase falls into the Luttinger-liquid (LL) class.\\
At half filling and $u\geq 0$, successive numerical investigations  have shown that the metal-insulator transition in fact moves to positive $u_c(x)$ values as soon as $x\gtrsim 1/2$,\cite{ADMO} reaching smoothly the $x=1$ value [$u_c(1)=4$]. Moreover, $\Delta_s$ turns out to be open for any $x\neq 1$ and $u<u_s(x)$. So that, at variance with the bosonization predictions at $n=1$, for $u<u_c(x)<u_s(x)$ the model behaves as a LEL, and a new phase appears, characterized by slowly decaying singlet-superconducting correlations and incommensurate modulations in the real-space charge correlations.\cite{AAA} The phase is denoted as ICSS. Moreover, for $u_c(x)\leq u\leq u_s(x)$, a fully gapped phase with bond ordered wave order is observed. The qualitative phase diagram of the model at half filling as derived in Refs. \onlinecite{ADMO} and \onlinecite{AAA} is given  in Fig. \ref{fig1}.

\subsection{Connection with the Simon-Aligia model}

One may wonder why the value $x=1/2$ plays a relevant role in the above results. In order to elucidate the issue, we observe that the properties of $H_{BC}$ --and in general of those Hamiltonians in which the interaction is local-- are better understood when these are represented in terms of on-site projection operators. The latter are defined as $X^{\alpha\beta}_i\doteq\ket{\alpha}_i\bra{\beta}_i$, where $\ket{\alpha}_i$
are the states allowed at a given site $i$, and $\alpha=\{0,\uparrow,\downarrow,2\}$, [$\ket{2}\equiv \ket{\uparrow\downarrow}$]. In such a language the nonvanishing entries of the Hamiltonian matrix representation are read directly as the nonvanishing coefficients of the projection operators. When rewritten in terms of these operators, $H_{BC}$ turns out to be a subcase of the more general Hamiltonian $H$ introduced by Simon and Aligia,\cite{SIAL}
\begin{widetext}
\be
    H= - \sum_{\langle ij\rangle, \sigma } \left [ X_i^{\sigma 0}X_j^{0\sigma}+ t_x X_i^{2\sigma}X_j^{\sigma 2}+ s_x \left ( X_i^{2 \sigma} X_j^{0 \bar \sigma}+X_i^{\sigma 0} X_j^{\bar \sigma 2} \right ) \right ] +  u \sum_i X_i^{22}
    \; , \label{ham_SA}
\ee
\end{widetext}
in which $t_x=(1-2 x)$ and $s_x=(1-x)$. Besides $u$, the behavior of $H_{BC}$ is determined by the strength of $t_x$ and $s_x$, since the original parameter $x$ appears in its matrix representation  only through these coefficients.  Importantly, $x=1/2$ implies $t_x=0$, whereas below and above such value $t_x$ changes sign. Unlike the overall sign of the hopping term, which can be easily adjusted by a particle-hole transform, the change in the sign of $t_x$ cannot be fixed. A negative $t_x$ induces frustration in the motion of pairs, which is also driven by  the positive $s_x$ term. In particular, for $|t_x|\gtrsim |s_x|$ --in our case $x\gtrsim 2/3$-- the mobility of the pairs should become favored in the system, at least for not too large $u$. Numerical results available so far refer instead mainly to the particle-hole symmetric choice $t_x=s_x$.\cite{NSA}

\subsection{$s_x=0$ case and phase separation}\label{secIIB}

The integrable instances of $H$ discussed in the literature are the standard Hubbard case ($t_x=1=s_x$), the infinite-$U$ Hubbard model ($t_x=0=s_x$), the $x=1$ subcase of $H_{BC}$ (Ref. \onlinecite{AAS}) ($t_x=-1$, $s_x=0$) and the model recently solved in Ref. \onlinecite{MONT} (arbitrary $t_x<0$, $s_x=0$), which could be representative of the regime $x\gtrsim 2/3$. In this case it is found that --whenever $u\leq u_c(x,n)$-- the ground state is characterized by PS in the range of filling values $n_l\leq n\leq n_h$. The two coexisting phases have different particle densities $n_l<1$ and $n_h>1$ amounting to two spinless-fermions (SF) systems. For this reason we shall denote the model as 2SF: in the low-density phase the SF fluid consists of $n_l$ electrons moving in a background of empty sites, whereas in the high-density phase it amounts to $2-n_h$ holes moving in a background of doubly occupied sites.  Both phases are in principle conducting, with incommensurate Fermi momenta $k_F^{(a)}= \pi n_a$ (where $a=l,h$). The two densities $n_l$ and $n_h$ can be expressed\cite{MONT} as functions of the parameters $u$, $t_x$, and $\mu$, as follows:
\begin{equation}
    n_l= 1-\frac{1}{\pi} \arccos \left (\frac{\mu}{2}\right ) \, , \; n_h= 1+ \frac{1}{\pi} \arccos \left (\frac{u-\mu}{2 |t_x|}\right ) \;,\label{neq}
\end{equation}
where $\mu$ is the chemical potential, which must satisfy the transcendental equation
\be
    \mu= \frac{1}{n_h-n_l} \left [ - {2\over\pi} |t_x|\sin {\pi n_h}+ u (n_h-1)+{2\over \pi} \sin {\pi n_l} \right ] \;. \label{mueq}
\ee

For $t_x=-1$ the $x=1$ subcase of $H_{BC}$ is recovered: in this case $n_h=2-n_l$.

The presence of PS at $x=1$ has a simple explanation. Let us assume for simplicity $u=0$. In the thermodynamic limit, the ground-state energy per site $E_0(1)$ coincides with that of $N_s$ SF ($N_s$ being the number of singly occupied sites) and is independent of the number $N_d$ ($N_e$) of doubly occupied (empty) sites:\cite{AAS} $E_0(1)=-2/\pi \sin(\pi N_s/L)$. As a function of the filling, the absolute minima in the two homogeneous phases (the one consisting of singly occupied and empty sites only, and the one consisting of singly and doubly occupied sites only) are reached at quarter filling ($n=1/2=n_s$) and three-quarter filling ($n=3/2$, $n_s=1/2$). At any $n$ within the range $n_l=1/2\leq n \leq 3/2=n_h$, one can use the Maxwell construction: $n_s$ assumes the value $1/2$, and $(2 n-1)/4$ doubly occupied sites are added to the ground state at zero-energy cost, i.e., we are in presence of PS. On the contrary, at $x=0$ the ground-state energy is that of a system of $n$ electrons with spin moving on a chain, $E_0(0)=-4/\pi \sin (\pi n/2)$: in this case the absolute minimum is reached at half filling ($n=1$) and no PS is observed. For continuity argument, one expects PS to take place also for $x$ sufficiently close to 1, and at positive, not too large, $u$.\\
The findings of Refs. \onlinecite{ADMO} and \onlinecite{AAA} at half filling and $x\gtrsim 1/2$ could be consistent with a scenario of PS. As an example, we note that within the 2SF model, the critical line for PS reads $u_c(x,1)=4 x$, which is quite close to the numerical transition line of Fig. \ref{fig1} in the vicinity of $x=1$. In order to verify our hypothesis, a thorough analysis of the regime $n\neq 1$ is required. The available numerical simulations in this case\cite{QSZ,AAG} address mainly the issues of spin gap opening and pair-pair correlations above half filling, which features one would expect from the bosonization approach at $x\ll 1$.  The focus of the next sections is instead on the ground-state energy, chemical potential, charge-charge and pair-pair correlations, in the whole range $n\neq 1$.

\section{Ground-state energy and phase separation\label{secIII}}

\begin{figure}
\includegraphics[width=80mm,keepaspectratio,clip]{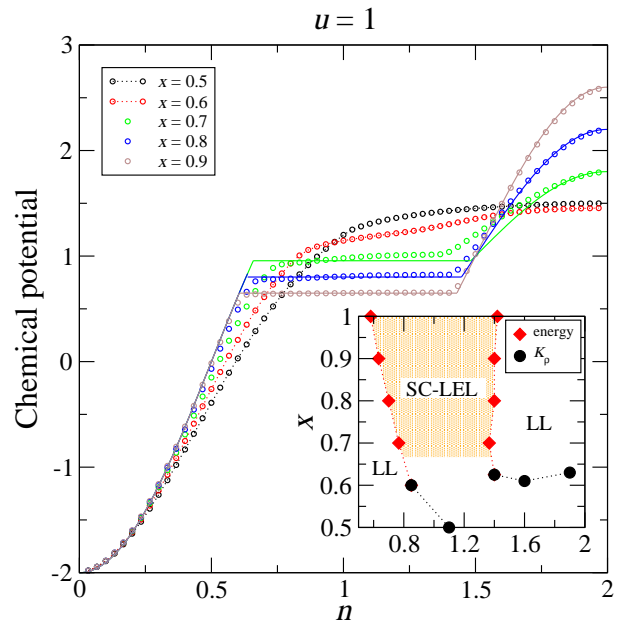}
\caption{(Color online) Discrete derivative $[E_0(n+\delta n)-E_0(n-\delta n)]/2\delta n$ of the DMRG ground-state energy $E_0$ at $u=1$: $x$ increases from top to bottom with reference to $n=1$. The continuous lines represent the exact results at $s_x=0$ (Ref. \onlinecite{MONT}). The numerical phase diagram is given in the inset: red diamonds are determined from the end points of the plateaux, and black dots refer to the analysis of $K_\rho$ specified in the text. In the SC-LEL phase the shaded region is characterized by NPS.}
\label{mu+xc_vs_n_u1.0-OBC+th}
\end{figure}

\begin{figure}
\includegraphics[width=80mm,keepaspectratio,clip]{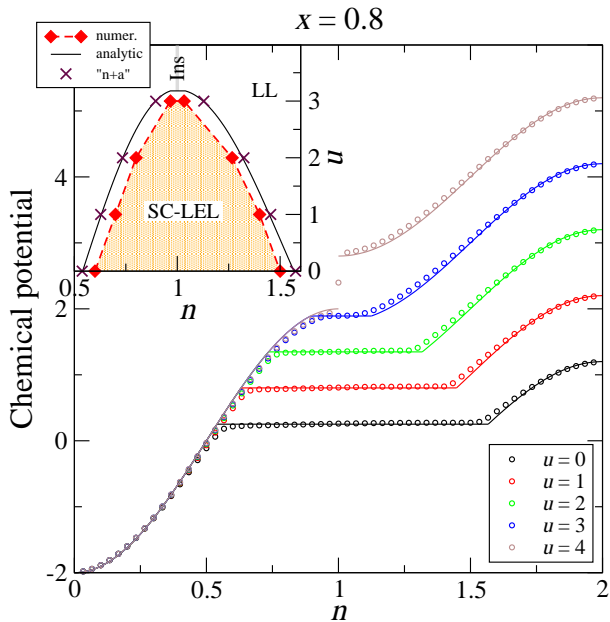}
\caption{(Color online) Same as Fig. \ref{mu+xc_vs_n_u1.0-OBC+th} but with fixed $x=0.8$ and varying $u$ (increasing from bottom to top). In the inset the continuous line is the analytical estimate of the phase diagram according to Sec. \ref{secIIB}, while the crosses ("n+a") mark the transition points obtained by intersecting the analytical curves out of the plateaux with the numerical values of the chemical potential within the plateau. The shaded region is characterized by NPS. The jump at $n=1$ and $u=4$ corresponds to the nonvanishing charge gap of the insulating phase.}
\label{mu+uc_vs_n_x0.8-OBC+th}
\end{figure}

Thanks to the fact that the total number of electrons $N$ is always a good quantum number, for every value of $u$ and $x$ we have computed numerically the ground-state energy density $E_0$ in the range $0 \le n \le 2$ by a series of different runs varying $N$ from 0 to $2L$ in steps of 2, so that it is possible to fix in each run also the total magnetization to 0. We have used open boundary conditions and up to 768 DMRG states with three finite-system sweeps in order to improve the accuracy. The finite-size dependence on $L$ has been investigated by means of preliminary runs at $L=10$ and $L=50$; we have checked that the essential quantitative features of the curves $E_0(n)$ and $\partial E_0 / \partial n $ (approximated with finite differences $\delta n = 2/L$) depend weakly on $L$. The interesting feature that we observe for suitable values of $u$ and $x$ is that there is a range of filling values $n_l \le n \le n_h$ where the energy density is linear as a function of $n$. This means that in the grand-canonical ensemble by selecting a suitable value of chemical potential one could change the number of particles with no energy cost. In Fig. \ref{mu+xc_vs_n_u1.0-OBC+th} we report the results for the ``canonical'' chemical potential $\partial E_0 / \partial n $ with $L=60$ sites and $u=1$ at different values of $x$, while in Fig. \ref{mu+uc_vs_n_x0.8-OBC+th} we fix $x=0.8$ and vary $u$ in the range $0-4$ again with $L=60$. In both cases we observe that there exists a flat region of values of $n$ corresponding to an infinite charge compressibility $\kappa$,
\begin{equation}
    \kappa^{-1} = n^2 \frac{\partial^2 E_0}{\partial n^2} .
\end{equation}
The size of this region increases with increasing $x$ and diminishing $u$, while for instance at $x=0.5$ and $u=1$ it is absent. For reasons which will become clear in the next sections, we denote such a region as SC-LEL. The numerical phase diagrams in the planes $(n,x)$ at $u=1$ and $(n,u)$ at $x=0.8$ are plotted in the insets of Figs. \ref{mu+xc_vs_n_u1.0-OBC+th} and \ref{mu+uc_vs_n_x0.8-OBC+th} and are obtained by tracing the (discrete) values of $n$ at which $\partial E_0/\partial n$ starts to be flat. We should observe that the energy density in the SC-LEL region does not show a convex shape so we may safely extract the transition lines from the end points of the plateaux.\cite{CSC} On the one hand, a finer inspection reveals that, due to numerical errors and/or finite-size effects, the edges of the plateaux show some rounding in certain cases. On the other hand, we can compare our numerical results with the transition lines obtained analytically in the exactly solvable model of Ref. \onlinecite{MONT}, where $s_x=0$, by taking here $t_x=1-2x$, and
\begin{equation}
    \frac{\partial E_0}{\partial n} = \left\{
    \begin{array}{lcl}
        -2 \cos({\pi n}) & \mbox{for} & n \le n_l \\
        u+2(2x-1) \cos({\pi n}) & \mbox{for} & n \ge n_h\;,
    \end{array}\right.
\end{equation}
where $n_l$ and $n_h$ together with the constant value of the chemical potential $\mu$ in the region $n_l\leq n\leq n_h$ are found by solving Eqs. (\ref{neq}) and (\ref{mueq}). The comparison between DMRG and analytical estimates of $n_l$ and $n_h$ in Fig. \ref{mu+uc_vs_n_x0.8-OBC+th} reveals a very good agreement at $x=0.8$, meaning that such a value is already representative of the physics of the exactly solvable point $s_x=0$. In fact, at $x=0.8$ one has $t_x=-0.6$ and $s_x=0.2$. The agreement diminishes for $x=0.7$ (Fig. \ref{mu+xc_vs_n_u1.0-OBC+th}) and for $x=0.6$ from our finite-size data ($L=60$) we are unable to discern a plateau of nonzero width. This latter fact is in accordance with the observation that at $x=0.7$ $s_x$ and $t_x$ are almost of the same order (namely, $s_x=0.3$ and $t_x=-0.4$), though still $-t_x>s_x$, whereas at $x=0.6$ $s_x$ is even greater than $|t_x|$.

The SC-LEL regions $x \geq x_c(u)$ at fixed $u$, or $u \leq u_c(n) $ at fixed $x$, are the candidates for the ICSS phase away from the half-filling situation, where the transition points are indeed consistent with our previous analysis of the ICSS region. In order to confirm this indication an analysis of the spin gap and of the pair-pair, charge-charge correlations is required.

\section{Correlations and spin gap vs filling\label{secIV}}

The study of the previous section produced evidence of the fact that the coexistence of phases at different densities characterizes the system's behavior in a wide range of filling values $n_l\leq n\leq n_h$, for $u\leq u_c(x,n)$. In particular, at $n=1$ it seems to be present at least for $x\gtrsim 2/3$ within the whole ICSS phase. Still at half filling it was noticed\cite{AAA} that, due to the presence of an open spin gap, both pair-pair
$P_r\doteq\langle c_{i\uparrow}^\dagger c_{i\downarrow}^\dagger c_{i+r\downarrow} c_{i+r\uparrow}\rangle -
\langle c_{i\uparrow}^\dagger c_{i+r\uparrow} \rangle \langle c_{i\downarrow}^\dagger c_{i+r\downarrow}\rangle$
and charge-charge $C_r\doteq\langle n_{i} n_{i+r}\rangle-\langle n_{i}\rangle \langle n_{i+r}\rangle$ (where $n_i=n_{i\uparrow}+n_{i\downarrow}$) correlations become dominant for $u\leq u_c(x,1)$.

\begin{figure}
\includegraphics[width=90mm,keepaspectratio,clip]{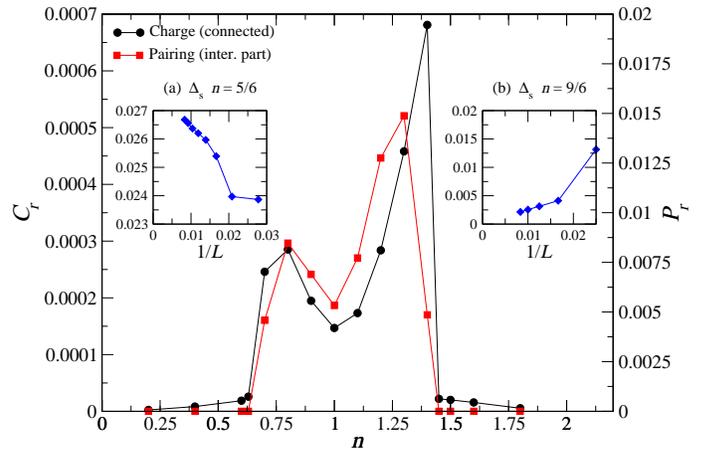}
\caption{(Color online) Charge-charge (black circles) and pair-pair (red squares) correlations between site $L/3$ and site $2L/3$ (open chains) versus filling $n$. Here $x=0.8$ and $u=1$. In the two insets we have reported the finite-size spin gaps computed as $\Delta_s=E_0(S_z=1)- E_0(S_z=0)$ ($S_z$ being the total $z$ component of the spin) for (a) $n=5/6 > n_l$ ($\Delta_s \neq 0$) and (b) $n=9/6 \gtrsim n_h$ ($\Delta_s \to 0$).} \label{etap+nn+sg_vs_n_u1.0x0.8-OBC}
\end{figure}

In this section we then explore the filling dependence of $P_r$, $C_r$, and $\Delta_s$. In Fig. \ref{etap+nn+sg_vs_n_u1.0x0.8-OBC} we report $P_{L/3}$ and $C_{L/3}$ (between site $L/3$ and site $2L/3$ in an open chain) for a given $L=120$ at $u=1$, $x=0.8$. Two aspects emerge quite clearly in the data shown. First, both charge-charge and pair-pair correlations keep a significant track of the undergoing transition to the PS regime, becoming appreciably different from zero only in the range $n_l\leq n\leq n_h$. Second, both $n_l$ and $n_h$ are indistinguishable, within our numerical precision, from the values which limit the coexistence of phases described in the previous section. For this reason we have used the same symbols. It is also worth noticing that $P_r$ and $C_r$ have the same kind of dependence on $n$, in particular exhibiting a minimum at $n=1$ and reaching their absolute maximum for $n\approx 1.3$.

According to the theory of Luttinger liquids $P_r$ and $C_r$ are expected to become dominant in presence of an open spin gap, since their power-law decay with $r$ is determined by a different exponent in the LL case ($\Delta_s=0=\Delta_c$) and in the LEL case ($\Delta_s\neq 0$). Hence, we may expect that $\Delta_s\neq 0$ in the same range of filling values in which $P_{L/3}$ and $C_{L/3}$ are. Accurate DMRG simulations show that this is in fact the case. The two insets of Fig. \ref{etap+nn+sg_vs_n_u1.0x0.8-OBC} report the finite-size scaling of $\Delta_s$ for two significant cases: (a) $n_l\leq n\leq 1$ and (b) $n>n_h$. In both cases the results are in agreement with the findings for $P_r$ and $C_r$, though they differ from standard bosonization predictions: as explained in Sec. \ref{secII}, within the latter approximation at $u\geq 0$ the spin gap is found to be closed for any $n\leq 1$ and open for any $n>1$.\\
The extension of the above analysis to values of $x$ within the range $1/2\lesssim x\lesssim 2/3$ shows how the results are qualitatively similar. They can be resumed in the observation that for $u\leq u_c(x,n)$ and for $n_l\leq n\leq n_h$ the system becomes a LEL, characterized by an open spin gap and dominant pair-pair and charge-charge correlations.
The observation can be exploited to infer the numerical values of $n_l$ and $n_h$ also in a region where the chemical potential does not show evidence of phase coexistence. As an example, from the behavior of $P_{L/3}$ (see inset of Fig. \ref{KrhoVSnX06U1}) we obtained $n_l$ at $x=0.6$ $u=1$ as reported in fig. \ref{mu+xc_vs_n_u1.0-OBC+th}.
Further information on the nature of the pairs in the LEL phase can be gained from the analysis of the pair structure factor $P(q)=\sum_r P_r \exp{(i q r)}$ (not shown), which turns out to be peaked for vanishing $q$. This fact ensures that the electrons forming the pairs have opposite momenta ($k$, $-k$). In order to establish whether the LEL phase displays dominant superconducting ($K_\rho>1$) or charge ($K_\rho<1$) correlations, the derivation and analysis of the Luttinger parameter $K_\rho$ are now due.

\section{Charge Structure factor and $K_\rho$\label{secV}}

In this section we deepen our analysis of the phase diagram of the model under investigation by numerically evaluating the Luttinger parameter $K_\rho$. For the half-filled case, the latter was estimated in Ref. \onlinecite{AAA} by fitting the long-distance behavior of the equal time charge-charge and pair-pair correlation functions according to their asymptotic behavior, as predicted by conformal field theory equations\cite{JAMU} for a spin-gapped phase
\begin{eqnarray}
    C_r &\sim& \frac{K_\rho}{(\pi r)^2} + A\frac{\cos(2k_F r)}{r^{K_\rho}}\;,\label{LEL_CFT}\\
    P_r &\sim&  r^{1/K_\rho}\;.
\end{eqnarray}
There, it was shown that in the ICSS phase the dominant correlations are the superconducting ones, i.e., $K_\rho>1$ (for $x=0.8$ and $u=1.0$, $K_\rho\approx 1.3$).

An estimate of $K_\rho$ can also be extracted from the study of the static structure factor $N(q)=\sum_{r}e^{iqr} C_r$, since the Fourier transform of the non-oscillating term of Eq. (\ref{LEL_CFT}) gives the expression
\be
    K_\rho=\frac{\pi}{q}N(q\rightarrow 0)\;. \label{Krho}
\ee
Here we exploit this method to characterize the (super)conducting behavior of our system in the LEL regime. In our DMRG calculations we simulated an open chain of length $L=120$, taking $j=L/2$ to reduce the finite-size effects due to the presence of the borders.

\begin{figure}
\includegraphics[width=80mm,keepaspectratio,clip]{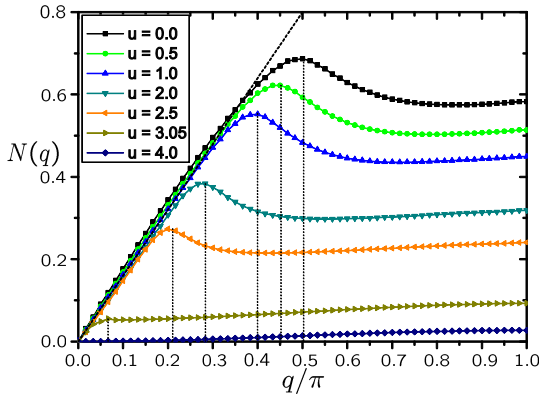}
\caption{(Color online) Static charge structure factor at half-filling for $x=0.8$ and various $u$. The slope of the dashed line is $1.6$. For each curve, dotted lines are drawn in correspondence with $q=2 \pi n_d$.} \label{fig:Nq-x0.8n1}
\end{figure}
\begin{figure}
\includegraphics[width=80mm,keepaspectratio,clip]{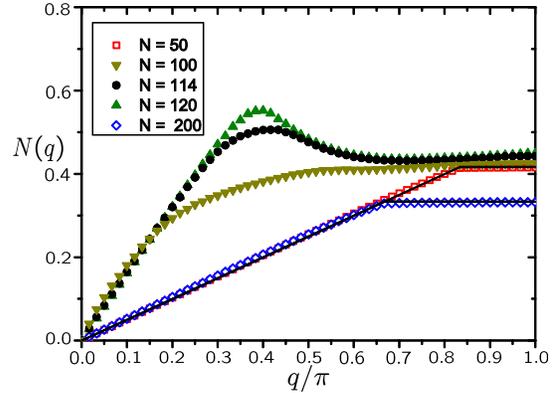}
\caption{(Color online) Static charge structure factor at $x=0.8$ and $u=1.0$ for several fillings $n$ ($L=120$). The continuous lines represent the analytical results obtained within the 2SF model in the low- and high-density regimes at the same $n$ and $t_x$ ($=0.6$) values.} \label{fig:Nq-X08U1-Nvari}
\end{figure}
In Fig. \ref{fig:Nq-x0.8n1} we plot $N(q)$ at $x=0.8$, $n=1$ and various $u>0$. One can see that within the ICSS phase, i.e., $u \lesssim u_c(0.8,1)=3.05$, the slope of the charge structure factor is weakly dependent on $u$, always implying a $K_\rho>1$: in this case $K_\rho\approx 1.6\pm 0.2$. The latter is an overestimation of the actual value of $K_\rho\approx 1.3$.\cite{AAA} This is in fact a general feature of the method employed since $N(q)$ is numerically obtained for a finite lattice and logarithmic correction should be included. Nevertheless the present estimation is consistent with the previous result.

Figure \ref{fig:Nq-x0.8n1} also shows that the maximum position $q^*/\pi$ depends on $u$. For a LEL we expect such a maximum to occur for $q^*=2 k_F$, where at half filling $k_F=\pi/2$.  Here instead the actual value of $q^*$ appears to be determined by the density  of doubly occupied sites $n_d$, as the dotted lines reported in Fig. \ref{fig:Nq-x0.8n1} in correspondence with the value $q=2 \pi n_d$ show. This feature suggests that in the ICSS phase the system behaves as an effective liquid of $n_d$ hard-core bosons, at least for large enough $x$ values.
\begin{figure}
\includegraphics[width=80mm,keepaspectratio,clip]{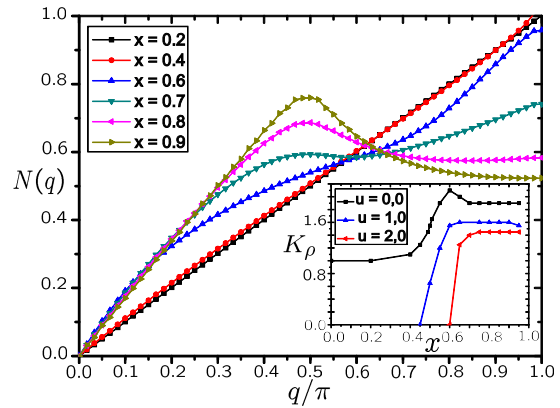}
\caption{(Color online) Static charge structure factor $N(q)$ at half-filling and $u=0.0$ for several $x$ values. Inset: $K_\rho$ derived from $N(q)$ as a function of $x$ for three different values of $u$. The lines are just guides for the eyes.} \label{fig:Nq-n1U10Xvari}
\end{figure}
The presence of a marked peak at $q=q^*$ is reduced away from the half-filling case, as can be seen in Fig. \ref{fig:Nq-X08U1-Nvari}, where one observes the different behavior of $N(q)$ as a function of the filling. Remarkably, in the low- and high-density regimes the results coincide with those obtained analytically for the 2SF model at the same $t_x$ value and $s_x=0$ ($K_\rho=1/2$ and $q^*/\pi=4 k_F$). Whereas, within the phase characterized by $K_\rho>1$, the height of the maximum decreases and the width increases moving from half filling, to disappear completely in proximity of $n_l$ and $n_h$. The maximum at $q^*\approx 2 \pi n_d$ disappears also at half filling by sufficiently lowering $x$. As shown in Fig. \ref{fig:Nq-n1U10Xvari}, for $0.5\lesssim x\lesssim 0.7$ it correctly moves to $q^*=\pi$, with $K_\rho>1$.

\begin{figure}
\centering
\includegraphics[width=80mm,keepaspectratio,clip]{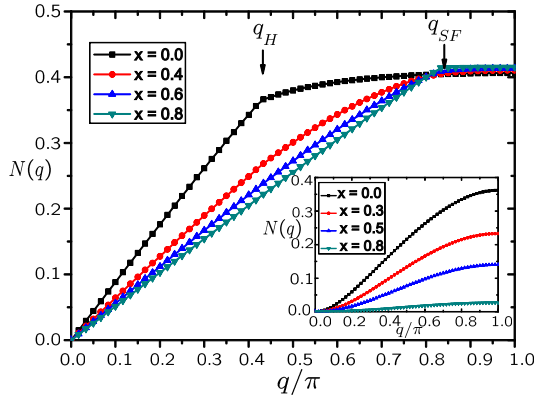}
\caption{(Color online) Static charge structure factor at filling $n=5/12$ and $u=1.0$ plotted for several values of $x$. Inset: static charge structure factor at half filling in the insulating regime ($u=4.0$) for several values of $x$.}\label{fig:Xvari}
\end{figure}

The effect of the bond-charge interaction $x$ on $N(q)$ is visible also outside the LEL regime. Figure \ref{fig:Xvari} shows how, by increasing $x$, there is a crossover both in the conducting regime at low density $n=5/12$ and in the insulating regime at half filling with strong interaction $u=4$ (inset). In the LL case it is seen that the dominant modulation in $N(q)$ moves from the Hubbard type case with spin, $q_H=2 k_F$ to the value in the spinless fermion case $q_{\mbox{SF}}=4 k_F$, $K_\rho$ varying correspondingly. In the insulating case (inset), one observes the crossover between antiferromagnetic ($x=0$, $u>0$) and the paramagnetic ($x=1$, $u\ge 4$) insulator: the charge structure factor is increasingly suppressed by enhancing the bond-charge parameter $x$. To summarize, in both cases increasing $x$ drives the system toward a SF picture.

\begin{figure}
\includegraphics[width=80mm,keepaspectratio,clip]{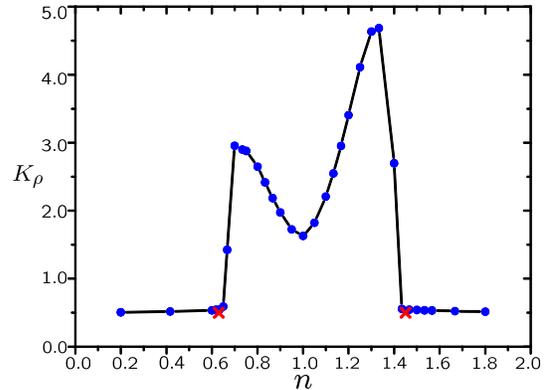}
\caption{(Color online) $K_\rho$ at $x=0.8$ and $u=1.0$ as a function of the filling $n$ ($L=120$). The line is just a guide for the eyes and the crosses mark $n_l$ and $n_h$ as determined in Sec. \ref{secII}.} \label{fig:Krho-X08U1-Nvari}
\end{figure}

\begin{figure}
\includegraphics[width=80mm,keepaspectratio,clip]{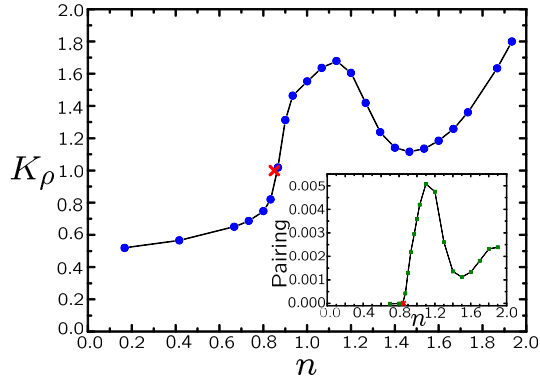}
\caption{(Color online) $K_\rho$ vs $n$ at $x=0.6$ and $u=1$. Inset: pair-pair correlations between site $L/3$ and site $2L/3$ (open chains) in same case.} \label{KrhoVSnX06U1}
\end{figure}

An exhaustive study of the static charge structure factor has been performed, so that by using Eq. (\ref{Krho}) we are able to follow the behavior of $K_\rho$ across the transition to the SC-LEL regime both at half filling (inset of Fig. \ref{fig:Nq-n1U10Xvari}) and with varying the filling (Fig. \ref{fig:Krho-X08U1-Nvari}). At $n=1$ the case $u=0$ is peculiar in that the SC phase is entered from a conducting phase (with $K_\rho=1$), while at any $0<u<u_c(x,n)$ the system is otherwise insulating ($K_\rho=0$).\\
The filling dependence of $K_\rho$ is shown in Fig. \ref{fig:Krho-X08U1-Nvari} at $x=0.8$. Remarkably, $K_\rho$ has the same qualitative behavior of pair-pair and charge-charge correlations (see Fig. \ref{etap+nn+sg_vs_n_u1.0x0.8-OBC}), in agreement with that expected from the conformal field theory equations (\ref{LEL_CFT}) for a LEL. In particular, it is seen that all the three quantities reach their maximum for $n\approx 1.3$. Moreover $K_\rho>1$ in the same region in which $P_{L/3}$ and $C_{L/3}$ are nonvanishing, the latter being very well approximated by the range $n_l\leq n\leq n_h$, with $n_l$ and $n_h$ as calculated for the 2SF model (same $t_x=0.6$, $s_x=0$).\\
The situation is slightly different at $x=0.6$. As shown in Fig. \ref{KrhoVSnX06U1}, both $K_\rho$ and $P_{L/3}$ again have the same qualitative behavior,  so that also in this case we can safely estimate $n_l$ and $n_h$ as the values of $n$ at which, for instance, $K_\rho$ becomes smaller than 1. In this way we obtained the points reported in the phase diagram $(x,n)$ also at $x=0.6$ (see Fig. \ref{mu+xc_vs_n_u1.0-OBC+th}), in which case the analysis of the ground-state energy was not conclusive about the presence of PS. Notice that the value $n_h=2$ at $x=0.6$ is substantially greater with respect to the one obtained at $x=0.7$ by the analysis of the chemical potential. This again could be a signal of the crossover to a regime in which the physics of the weak coupling limit ($x\ll 1$) begins to emerge, merging with that of the strong coupling case $x\simeq 1$. In fact, $n_l<1$ is characteristic of the latter regime. In this case both $K_\rho$ and $P_{L/3}$ reach their maximum value for $n\approx 1.1$.

Finally, at $x=0.5$ and $u=1$, previous numerical analyses\cite{KESC, AAG, ADMO,AAA} suggest that the system is already in the weak coupling limit. The point in the inset of Fig. \ref{mu+xc_vs_n_u1.0-OBC+th} for this value is simply obtained by imposing $K_\rho=1$ in Eq. (\ref{ubos}).

\section{Discussion and conclusions\label{secVI}}

The analyses of the previous paragraphs concur to complete the scenario of the physics described by the Hirsch model at $u\geq 0$, arbitrary filling $n$, and $0\leq x\leq 1$. Depending on the value of the bond-charge interaction $x$, one can distinguish three regions. The results are in fact representative of those relative to the more general Simon-Aligia Hamiltonian (\ref{ham_SA}) and can be resumed as follows:
\begin{itemize}
\item{}For $0\leq x\lesssim 1/2$ ($t_x>0$) the system behaves as a Luttinger Liquid (LL) for $n<1$, as a Mott insulator for $n=1$, and possibly there is a transition to a superconducting (SC) Luther-Emery liquid (LEL) phase for $n>1$, in agreement with bosonization predictions.
\item{}For $2/3\lesssim x\leq 1$ ($t_x\lesssim -s_x\lesssim 0$)  one can distinguish a high- ($n\geq n_h$) and a low-density ($n\leq n_l\le 1$) region of spinless fermion (SF) LL. These coexist up to $u_c(x,n)$ in the range $n_l\leq n\leq n_h<2$ in a phase of nanoscale phase separation (NPS, see below) with SC properties. Within the NPS phase the LEL manifests  an incommensurate modulation related to the number $n_d$ of doubly occupied sites.
\item{} For  $1/2\lesssim x\lesssim 2/3$ ($-s_x\lesssim t_x\lesssim 0$) we are in an intermediate regime. A homogeneous LEL phase with dominant SC correlations appears for $u\leq u_c(x,n)$ in a wider range $n_l\leq n\leq 2$ ($n_l\leq 1$).
\end{itemize}
Equation (\ref{ubos}) for $K_\rho$ at $x\ll 1$ suggests the mechanism driving the transition to the SC-LEL state for $n\leq 1$:  $K_\rho$ in this case is a decreasing function of $x$, so that the electrons progressively tend to behave as spinless fermions (SF), which feature holds exactly when $K_\rho=1/2$. At sufficiently low densities the change induced by increasing $x$  is seen as a crossover in the static charge structure factor $N(q)$: the cusp moves from $q=2k_F$ to $q=4 k_F$. Also, at $n=1$ and for large enough Coulomb repulsion $u>4$, the change appears as a crossover from an antiferromagnetic to a paramagnetic insulator. The acquired SF nature implies that the energy as a function of the filling reaches its minimum value at $n<n_l<1$, so that for greater filling values a regime of PS is favored. This possibility was already proven to work exactly at $s_x=0$n (Ref. \onlinecite{MONT}) ($x=1$), in which case also for $n>n_h$ the system behaves as a SF fluid, and the regime of PS amounts to the coexistence of the high- and the low-density fluids.

The  analytical results obtained in that case are in good quantitative agreement with some aspects of the results reported here, at least for $x\gtrsim 2/3$. Nevertheless, in order to acquire SC properties, one central feature was missing in the exact case: the presence of the spin degrees of freedom.  The inclusion of the latter allows for the opening of the spin gap: exactly at $s_x=0$ the ground state is fully degenerate with respect to the spin orientation, so that the spin-gap amplitude is vanishing. As soon as $s_x\neq 0$, the spins rearrange at short distance, and the PS state further lowers its energy by opening a spin gap. This phenomenon induces a shorter scale in the size of the coexisting phases. While in absence of the spin degrees of freedom the low- and high-density SF fluids would be spatially segregated; when spin is considered, the short-distance relative orientation of the electrons spin becomes relevant. The coexisting phases split into droplets of nanoscale size $\lambda$ entering the NPS phase: an incommensurate modulation appears in the liquid related to the number of droplets. This is determined by the number of doubly occupied (empty) sites $N_d$ ($N_e$): explicitly, $ L/N_e\lesssim \lambda\lesssim L/N_d$. The effect is particularly evident at half filling, where $N_d=N_e$ and the charge structure factor exhibits a neat maximum at $q=2 \pi N_d/L$ for $x\geq 2/3$. The maximum moves to $q=\pi$ for $1/2\leq x\leq 2/3$. This fact can be interpreted as a crossover of the system to a homogeneous phase, where the spin gap is open, SC correlations are dominant, and there is no longer a length scale determined by the size of the droplets. As a further confirmation, in the same intermediate region the maximum in $N(q)$ is observed at $q=\pi (2-n)$ for $n>1$, consistently with the  SC-LEL phase already present in this case at $x\leq 1/2$. The existence of droplets for $x\geq 2/3$ is also supported by the observation that the correct behavior of $N(q)$, which is in principle achieved by summing $C_r$ over all the range of $r$ values, can already be obtained, in its essential features, limiting the sum to the first $\lambda$ neighbors (not shown).

One recognizes {\it a posteriori} that increasing the bond-charge interaction $x$ from $0$ to $1$ in $H_{BC}$ amounts to passing, in the Simon-Aligia model of Eq. (\ref{ham_SA}), from the universality class of the Hubbard model ($t_x=s_x=1$) to that of the 2SF model discussed in Ref. \onlinecite{MONT} (arbitrary $t_x\le 0$, $s_x=0$). It would be interesting to exploit this observation within the bosonization approach, complementing the results for the two fluids of electrons with opposite spin at $x \ll 1$ (Ref. \onlinecite{JAMU}) with a bosonization study of the 2SF model, in the limit of weak coupling $s_x$ between the two SF fluids.

We also notice that within the NPS phase, for $u\leq u_c(x,n)$ the actual value of $u$ fixes $n_d$ and hence the size of the droplets. The analysis at $s_x=0$ shows that the mechanism survives and even enriches also for $u\leq 0$, and we expect further interesting physics to emerge also in relation to the context of cold fermionic atoms trapped in optical lattices. In this respect, it has been recently proven\cite{DUAN} that the Simon-Aligia Hamiltonian (\ref{ham_SA}) is the correct candidate to describe these systems in proximity of a broad Feshbach resonance.

\begin{acknowledgments}
We are especially grateful to Fabio Ortolani for the DMRG code. We also thank Anders W. Sandvik and George I. Japaridze for useful comments. This work was partially supported by the Italian MIUR through the PRIN Grant No. 2007JHLPEZ.
\end{acknowledgments}

\end{document}